\PassOptionsToPackage{bookmarks=false}{hyperref}
\documentclass[sigconf]{acmart}
\usepackage{multirow}
\usepackage{booktabs}
\usepackage{caption} 
\usepackage{subcaption}
\usepackage[colorinlistoftodos,prependcaption,textsize=tiny]{todonotes}
\usepackage{listings} 
\usepackage{url}
\usepackage{algorithm}
\usepackage{algpseudocode}
\usepackage{times}
\usepackage{balance}
\usepackage{enumerate}

\floatname{algorithm}{Algorithm}

\copyrightyear{2018} 
\acmYear{2018} 
\setcopyright{acmcopyright}
\acmConference[SEAMS '18]{SEAMS '18: 13th International Symposium on Software Engineering for Adaptive and Self-Managing Systems }{May 28--29, 2018}{Gothenburg, Sweden}
\acmBooktitle{SEAMS '18: SEAMS '18: 13th International Symposium on Software Engineering for Adaptive and Self-Managing Systems , May 28--29, 2018, Gothenburg, Sweden}
\acmPrice{15.00}
\acmDOI{10.1145/3194133.3194147}
\acmISBN{978-1-4503-5715-9/18/05}

\fancypagestyle{arxiv}{\fancyhf{}\fancyfoot[C]{~\\~\\This is the authors' version of the work. It is posted here for your personal use. Not for redistribution. The Definitive Version of Record was published in SEAMS'18, May 28--29, 2018, Gothenburg, Sweden. DOI: \href{https://doi.org/10.1145/3194133.3194147}{10.1145/3194133.3194147}.}}

\begin{document}
\title{A Learning Approach to Enhance Assurances for Real-Time Self-Adaptive Systems}
\renewcommand{\shorttitle}{}

\author{Arthur Rodrigues, \\Ricardo Diniz Caldas,\\  Gena\'{i}na Nunes Rodrigues}
\orcid{-}
\affiliation{%
  \institution{University of Bras{\'{i}}lia}
  \city{Bras{\'{i}}lia} 
  \state{DF} 
  \country{Brazil}
}
\email{arthy.rf@gmail.com, rdinizcal@gmail.com, genaina@unb.br}

\author{Thomas Vogel}
\affiliation{%
  \institution{Humboldt-Universit\"{a}t zu Berlin}
  \city{Berlin} 
  \country{Germany}
}\email{thomas.vogel@cs.hu-berlin.de}

\author{Patrizio Pelliccione} 
\affiliation{%
 \institution{Chalmers $|$ University of Gothenburg}
 \city{Gothenburg} 
 \country{Sweden}}
 \email{patrizio.pelliccione@gu.se}
 
\renewcommand{\shortauthors}{}

\begin{abstract}
The assurance of real-time properties is prone to context variability. Providing such assurance at design time would require to check all the possible context and system variations or to predict which one will be actually used. Both cases are not viable in practice since there are too many possibilities to foresee. Moreover, the knowledge required to fully provide the assurance for self-adaptive systems is only available at runtime and therefore difficult to predict at early development stages. Despite all the efforts on assurances for self-adaptive systems at design or runtime, there is still a gap on verifying and validating real-time constraints accounting for context variability. To fill this gap, we propose a method to provide assurance of self-adaptive systems, at design- and runtime, with special focus on real-time constraints. We combine off-line requirements elicitation and model checking with on-line data collection and data mining to guarantee the system's goals, both functional and non-functional, with fine tuning of the adaptation policies towards the optimization of quality attributes. We experimentally evaluate our method on a simulated prototype of a Body Sensor Network system (BSN) implemented in OpenDaVINCI. The results of the validation are promising and show that our method is effective in providing evidence that support the provision of assurance. 
\end{abstract}


\maketitle
\thispagestyle{arxiv}

\section{INTRODUCTION} 
Self-adaptive systems (SAS) shall adapt to changing contexts conditions to meet functional and non-functional requirements~\cite{Lemos17SEfSAS}. For safety-critical systems such as medical applications, the performance and time constraints are domain requirements whose violation could lead to catastrophic failures, putting the user's life at risk. Therefore, to guarantee the safe operation of this kind of system, a model checking approach is often adopted to provide the goals' reachability values for every modeled path. Nevertheless, the uncertainty coming from the difficulty in predicting which context conditions the system will encounter, combined with the huge amount of possible configurations of a complex system, hinders the development of adequate adaptation policies at design time~\cite{taxonomy}. 

Through the analysis of runtime data obtained either by monitoring or simulating the SAS, \emph{learning approaches} can be used to cope with this limitation by identifying the active context and modifying the adaptation space at runtime~\cite{AlessiaACon}. Thus, context changes and user profiles that were not anticipated at design time are addressed by learning new adaptation rules dynamically, or by modifying and improving existing rules~\cite{ShariflooMQBP16}. However, the generation of all possible situations, exclusively at runtime, poses a risk to the system's performance, reliability, and real-time constraints; this indicates the need of a comprehensive design-time assessment combined with an efficient runtime monitoring and validation. 

Although there are some studies on assurance provision for SAS, their verification techniques that infer whether a software system complies with its requirements still neglect the context variability factor and the corresponding impact on real-time properties. Basically, the state-of-the-art approaches can be divided into three major categories~\cite{Lemos17SEfSAS}: human driven (e.g., formal proof), system-driven (e.g., runtime verification), and hybrid (e.g., model checking). Despite the ability of hybrid methods to provide assurance at both on-line and off-line stages, there is a certain unpredictability concerning the influence of context variability on real-time properties that still needs to be mitigated. As a result, such unpredictability may hinder altogether the required management of \emph{``the continuous collection, analysis and synthesis of evidence that will form the core of the arguments that substantiate the provision of assurances''}~\cite{Lemos17SEfSAS}.

Our proposed work aims at supporting the provision of such evidence with special focus on real-time properties of SAS. The proposal relies on contextual goal modeling, taking goals as first-class citizen of the self-adaptation, and follows the MAPE-K reference model for SAS. Initially, the elicited requirements and domain knowledge are carried in a Contextual Goal Model~(CGM). Before implementing the SAS, we build a formal model of its behavior and verify the real-time properties with UPPAAL, a modeling and verification tool for real-time systems~\cite{bllpw:dimacs95}, to create the foundation upon which the SAS will be constructed. After the verification, we run a data-mining process on runtime data obtained from the SAS. The process applies a \emph{transductive transfer learning} setting~\cite{TransferLearningSurvey} to create prediction models in a faster and cheaper way. Afterwards, we use classifiers such as decision trees~\cite{Quinlan} to detect hidden context correlations that impact the system dependability, specially w.r.t. real-time constraints. 
By these means, we combine off-line requirements elicitation and model checking with on-line data collection and data mining to provide evidence that supports the creation of methods to guarantee the system goals fulfillment. From the real-time perspective, in particular, adaptation plans that reconfigure the system can be developed so that the real-time properties are met. In case of adverse context conditions, our method suggests alternative plans capable of fulfilling such goals based on the decision tree obtained through the data mining process while respecting the real-time properties verified by the model checking process. By these means, we postulate that our method helps to build \emph{``the core of the arguments that substantiate the provision of assurances''} combining model verification and data mining in the continuous feedback loop of collection, analysis and synthesis of evidence. Additionally, the knowledge obtained by data mining guides the optimization of a target quality attribute through the refinement of the adaptation policies in a way to minimize the possible collateral effect.

We experimentally evaluate our method on a simulated prototype of a Body Sensor Network system (BSN)~\cite{Pessoa201754}. The experiment consists in analyzing the reachability of BSN goals by merging two different perspectives. The first one relies on the UPPAAL model-checking technique over the formal model of the BSN to extract the goals' reachability values. Afterwards, the verified model of the BSN is implemented in OpenDaVINCI~\cite{ODV}, a well-established distributed real-time platform, to simulate the BSN as a real-time SAS. The second one is based on the application of data-mining techniques, specifically classifiers, on the dataset obtained by simulation to discover how the system behaves in the presence of different contexts, particularly with respect to the real-time constraints of the BSN. The data generated by the simulated BSN is continuously stored in a database after each scheduling period and analyzed by the data-mining and learning process we propose. 

The learning process has been shown to be useful in raising the system's awareness towards operational contexts that might pose a threat at runtime to the real-time properties assured by the UPPAAL model checking. Therefore, it assists the development of an appropriate set of adaptation policies for the controller. In addition, the evaluation allowed us to simulate scenarios under varying numbers of active sensors, different modes of the controller, and varying health risks of patients. Moreover, the method has shown itself effective to provide optimization strategies for dynamic adaptation of the controller mode under adverse sensors battery conditions, which could be crucial for patients under critical health risk, still in accordance with the real-time properties verified off-line by model~checking.

The rest of the paper is organized as follows. Section~\ref{sec:BSN} provides a brief description of our running example. We present our methodology in Section~\ref{sec:Proposal} and evaluate our proposal through experiments in OpenDaVINCI in Section~\ref{sec:Experiments}. Section~\ref{sec:RelatedWork} highlights major related work. Finally, Section~\ref{sec:Conclusion} concludes along with future work.%
\section{\textbf{Example: Body Sensor Network}} \label{sec:BSN}

To discuss our proposed methodology, we use the example of a Body Sensor Network~(BSN)~\cite{Pessoa201754} throughout this paper. The main objective of the BSN is to keep track of a patient's health status, continuously classifying it into \emph{low}, \emph{moderate}, or \emph{high} risk and, in the case of any anomaly, to send an emergency signal to a central unit. The structure of the BSN is as follows: a few wireless sensors are connected to a person to monitor her vital signs, namely, an electrocardiogram sensor (ECG) for heart rate and electrocardiogram curve, an oximeter (SPO2) for blood oxidation and blood oxidation curve, and a temperature sensor (TEMP). Additionally, there may be a central node (Control Sensor) responsible for preprocessing the collected data, filtering redundancy, or translating communication protocols. Table~\ref{operat} shows how the sensor values (and thus the context) relate to the patient's health risk as specified by a domain expert.

\begin{table}[htb]
\caption{Context operationalization for patient's status.}
\label{operat}
\small
\begin{tabular}{ll}
\multicolumn{1}{c}{\textbf{Sensor Information}} & \multicolumn{1}{c}{\textbf{Data Ranges}}                     \\\hline
Oxygenation:      & 100 $>$ low $>$ 94 $>$ moderate $>$ 90 $>$ high $>$ 0                        \\\hline
Pulse Rate:       & high $>$ 120 $>$ low $>$ 80 $>$ high $>$ 0                                 \\\hline
Temperature:     & \begin{tabular}[c]{@{}l@{}}50 $>$ high $>$ 38 $>$ moderate $>$ 37 $>$ low $>$ \\
 35 $>$ moderate $>$ 30 $>$ high $>$ 0\end{tabular} \\\hline
\end{tabular}
\end{table}

\subsection{A Contextual Goal Model of the BSN}

\begin{figure}[htb]
    \centering
    \includegraphics[width=\columnwidth]{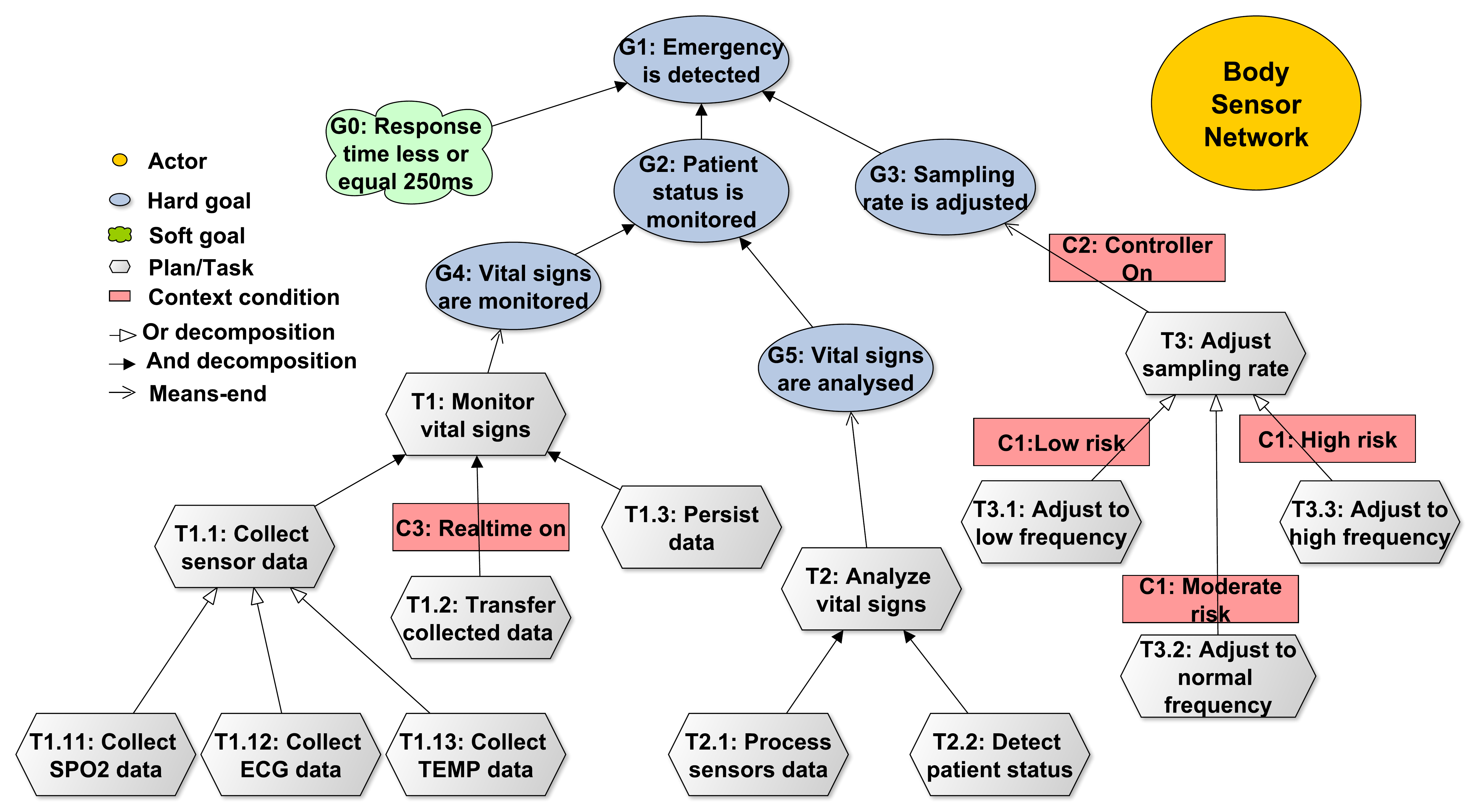}
    \caption{Contextual Goal Model of the BSN.}
    \label{fig:goaltree}
\end{figure}

Modeling a SAS requires to take into consideration not only the requirements and means to achieve them, but also the contextual information that may be related to the system's operation. For this purpose, we use a Contextual Goal Model~(CGM) since it allows us to specify in a simple structure the stakeholders and high-level requirements, the ways to meet such requirements, and the environmental factors that can affect the quality and behavior of a system. Figure~\ref{fig:goaltree} shows an excerpt of the CGM for the BSN.

According to Ali et al.~\cite{Ali}, a CGM is composed of: 
(i) actors such as humans or software that have goals and can decide autonomously on how to achieve these goals; 
(ii) goals as a useful abstraction to represent stakeholders' needs and expectations, offering an intuitive way to elicit and analyze requirements; 
(iii) tasks as atomic parts that are responsible for the operationalization of a system goal, that is, an operational means to satisfy stakeholders' needs; and 
(iv) contexts as partial states of the world that are relevant to a goal. 
A context is strongly related to goals since context changes may affect the goals of a stakeholder and the possible ways to satisfy the goals. 
Goals and tasks of a CGM can be refined into AND-decomposition (OR-decomposition), that is, a link that decomposes a goal/task into sub-goals/tasks, meaning that all (at least one) of the decomposed goals/tasks must be fulfilled/executed to satisfy its parent entity. 
The link between a goal and a task is called means-end, and indicates a means to fulfill a goal through the execution of a task.

According to Figure \ref{fig:goaltree}, the root goal of the actor BSN is ``G1: Emergency is detected''. G1 is refined into ``G2: Patient status is monitored'' and ``G3: Sampling rate is adjusted'' by an AND decomposition. G2 is divided into two subgoals: ``G4: Vital signs are monitored'' and ``G5: Vital signs are analysed''. Such goals are decomposed, within the boundary of the BSN actor, to finally reach executable tasks. 
The operation of the BSN is subject to three different context conditions. The first context (C1) is the aforementioned patient status, which may assume three possible values: low, moderate, and high risk. The second context (C2) is related to the controller's mode, which may be \emph{on} or \emph{off} and adapts the sampling rate of the patients' vitals depending on C1. The last context (C3) concerns the real-time mode of the sensor nodes' scheduler that can be \emph{on} or \emph{off}. Thus, C2 and C3 determine whether the BSN has activated or not the controlled mode respectively the real-time mode.

Each combination of executable tasks might contain different conjunction of contexts, and each conjunction (i.e., context of a goal model variant~\cite{Ali}) shapes the system to fulfill a requirement at a different quality level. For this work, we use the three context conditions modeled in the CGM (Figure~\ref{fig:goaltree}) as they impact the real-time properties (C3), quality (C2) and dependability attributes (C1).

\subsection{BSN Architecture}

\begin{figure}[!htb]
	\centering
	\includegraphics[width=0.65\columnwidth]{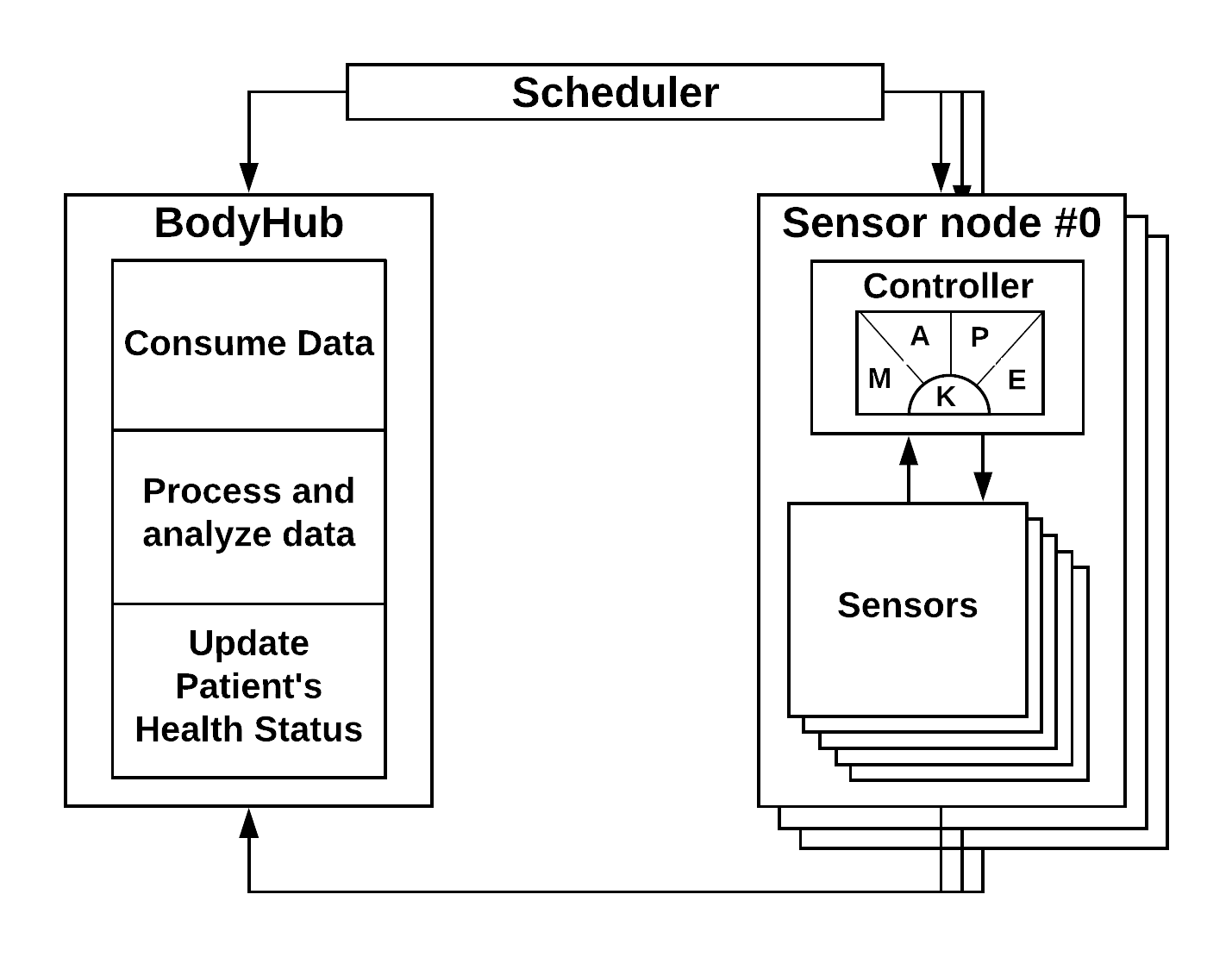}
	\caption{Architecture of the BSN.}
	\label{fig:archimpl}
\end{figure}

The BSN requires a network of distributed devices responsible for the execution of the tasks defined in the CGM (see Figure~\ref{fig:goaltree}). This network is defined by the architecture in Figure~\ref{fig:archimpl} and consists of (i) a Scheduler, (ii) a set of Sensor Nodes, and (iii) a BodyHub. 

The scheduler realizes the deterministic execution of the other modules by dictating their execution sequence using a first-come first-served (FCFS) algorithm. Thus, the scheduler exclusively commands the BodyHub or an active sensor node to execute by sending fixed period release signals. 
Each sensor node is a self-adaptive device capable of capturing sensor signals, processing, storing, and eventually sending data through wireless communication. A wide range of configurations may be applied regarding its components to fulfill the requirements based on policies (e.g., if battery $\leq 50\%$ then activate \emph{controlled mode}). Each sensor node operationalizes the tasks T1.1 (Collect sensor data) and T3 (Adjust the sampling rate) (cf.~Figure~\ref{fig:goaltree}), where the second one encompasses the following self-adaptive behavior when the controlled mode is activated: each node \emph{monitors} its sensor data, \emph{analyzes} the data to determine the patient's health risk, \emph{plans} a new sampling frequency based on the analysis, and finally \emph{executes} the needed change. Changing the sampling frequency of a sensor node influences the reliability of the sensed data and the battery consumption of this node. 

The BodyHub acts as an information centralizer for the data provided by the sensor nodes. It consumes, processes (e.g., store and fuse), and analyzes the data to decide and update the overall health status of the patient. Thus, it operationalizes the tasks T1.3 (Persist data) and T2 (Analyze vital signs) with its refinements (cf.~Figure~\ref{fig:goaltree}).
\section{OUR APPROACH FOR ASSURANCE OF REAL-TIME SAS}
\label{sec:Proposal}

The software engineering process for building self-adaptive safety-critical systems (e.g., self-adaptive real-time systems) must follow a guideline with perpetual assurances of goals from design to run-time. Knowing that feedback loops supported by processes should provide the basis for managing, among other things, the continuous synthesis of evidence~\cite{Lemos17SEfSAS}, we propose a feedback loop that combines off-line requirements elicitation and model checking with on-line data collection and data mining to provide the information that subsidize the provision of assurances. Hence, a means to guarantee the system goals fulfillment, both functional and non-functional, is using the knowledge obtained by the data mining to fine tune the adaptation policies towards the optimization of dependability attributes such as reliability, availability, or safety, as well as quality attributes like performance or energy-efficiency. The role of the data mining process is to discover and quantify the impact of operational contexts on predefined properties such as time constraints and quality attributes. Thus, adaptation policies can be developed to reconfigure the system in a way that respects the system's properties independently of the changing environment. 
To structure this idea, we describe our method as an enhanced feedback loop in the following steps:

\begin{enumerate}[i]
\item We start the method with a CGM as the specification of the stakeholders' needs (see Figure~\ref{fig:goaltree});

\item The next step is to model the core SAS architecture elements as well as their behavior and conduct a model-checking process with UPPAAL to verify the correctness of the SAS, especially whether real-time properties are satisfied or not. Then, we  devise the expected behavior of the SAS based on the analyzed context conditions;

\item After the verification stage, we implement the SAS and apply the concepts of transfer learning for the on-line assurance, in the sense that we learn from a simulation aiming at transferring the obtained knowledge to a real-world application. At this stage, we are concerned with the prototype implementation and its compliance with the properties verified by model checking while the next steps provide the on-line assurance;

\item We execute the prototype and collect the related runtime data such as the system's resource consumption and the occurring contexts conditions. The runtime data is stored as snapshots of the SAS, that is, we collect all the relevant variables and their respective values in progressive moments along the execution;

\item In possession of the collected runtime data, we apply a set of data mining algorithms aiming at identifying hidden correlations between the system's variables and the contexts and therefore, between combinations of contexts and the satisfaction of the system's properties;

\item Closing the feedback loop, the learning mechanism supports the provision of assurance by allowing the fine tuning of the (or suggesting novel) adaptation policies taking runtime data into account that was not anticipated when initially developing the policies. In case of new policies, a refactoring of the verified model and system is performed to make sure the once assured properties still hold after the refactorings.
\end{enumerate}

Next, we present such steps in further details.

\subsection{From Goals to Model Checking}

In our approach, the SAS is first modeled by taking into consideration the CGM leaf-tasks operationalization (see Figure~\ref{fig:goaltree}) in conformity with the BSN architecture (see Figure~\ref{fig:archimpl}). Each module of the architecture is then modeled as a timed automaton in UPPAAL~\cite{Behrmann06atutorial}, where each automaton represents a module template which may contain one or more instances (e.g., multiple sensor nodes use the same behavior template). In the BSN, the modules follow a First-Come-First-Served (FCFS) scheduling behavior fulfilling a basic life cycle represented by locations named according to the progress of the module behavior: \emph{wait}, \emph{run} and \emph{idle} in which the \emph{run} location is constrained by a clock and characterizes each architectural module. When the scheduling cycle is finished, it sets the guard condition \emph{done} to true. Thus, our modeling strategy is to represent the life cycle of modules that denotes the progress of the modules' behavior by different locations and guard conditions in the UPPAAL model for the verification of reachability properties. 

For the sake of space, we provide details of the UPPAAL model for the BSN at GitHub\footnote{\url{https://github.com/rdinizcal/SEAMS18/tree/master/uppaal}}. In the next subsection, we show how we map the goals of a SAS into properties for the UPPAAL models using our running example.

\subsubsection{Properties Verification}

We specify the properties of the model to be verified in Timed Computational Tree Logic (TCTL)~\cite{Henzinger1994}, since its the UPPAAL language to verify the real-time properties of the formalized model. TCTL is a real-time variant of CTL aimed to express properties of timed automata. Like CTL, the model verification relies on state or path expressions regarding properties such as \emph{reachability}, \emph{safety}, or \emph{liveness}. TCTL extends CTL with atomic clock constraints over the clocks, typically the set of clocks in the timed automaton under consideration. The TCTL model-checking problem is to check for a given timed automaton TA and TCTL formula $\phi$ whether TA $\vDash$ $\phi$~\cite{Baier:2008:PMC}. In UPPAAL, the properties are specified with a subset of TCTL augmented with syntactical symbols such as $\to$ and logical operators like $imply$ and $>,\geq,<,\leq$ to include variables and time evaluations. Table~\ref{tab:TCTL-UPPAAL} lists such basic expressions.

\begin{table}[!htb]
    \caption{Basic TCTL expressions in UPPAAL.}
	\label{tab:TCTL-UPPAAL}
    \centering
    \small
    \begin{tabular}{cl}
        \hline
        \textbf{Expression} & \textbf{Semantics} \\ \hline
        E$\lozenge$ $\phi$ & there exists a path where $\phi$ eventually holds \\ \hline
        A$\square$ $\phi$ & for all paths $\phi$ always holds \\ \hline
        E$\square$ $\phi$ & there exists a path where $\phi$ always holds \\ \hline
        A$\lozenge$ $\phi$ & for all paths $\phi$ will eventually hold \\ \hline
        $\phi$ $\to$ $\psi$ & whenever $\phi$ holds $\psi$ will eventually hold \\ \hline
        $\phi$ imply $c \leq T$ & $\phi$ holds if and only if the clock $c$ is within T \\ \hline
        A$\square$ not deadlock & checks for deadlocks \\ \hline
    \end{tabular}
    \normalsize
\end{table}
\small
\begin{table*}[htb]
	\caption{Major TCTL properties for model checking the BSN goals.}
	\label{tab:properties}
    \centering
    \begin{tabular}{ccll}
    \hline
    \textbf{Goal} & \textbf{ID} & \textbf{Informal Description} & \textbf{Specification in TCTL} \\ \hline
    \multirow{3}{*}{\textbf{N/A}}
                                & &                                                             & \\                
                                & \textbf{P1} & The controlled system is \bf{deadlock free}.                     & $A\square \, not \, deadlock$ \\
                              & &                                                             & \\ \hline
\multirow{3}{*}{\textbf{N/A}}
                                & & Whenever the \bf{scheduler cycle} is completed,                     & $A\square \, scheduler.done \,$ imply $\, (bodyhub.done \, \&$ \\
                                & \textbf{P2} & implies that the bodyhub and the three                      & $sensornode(1).done \, \& \, sensornode(2).done \, \&$ \\
                                & & sensors have been executed.                                  & $sensornode(3).done)$ \\ \hline
    \multirow{5}{*}{\textbf{G1}}
                                & & Whenever the patients' health status is on high risk and an & \\
                                & & emergency has been detected it implies that the observer's clock& $A\square \,bodyhub.hstatus==high \, \& \, emergency==true$ \\
                                & \textbf{P3} & is less or equal 250 (ms) and a single scheduler  & imply $\, observer.o\_clk \leq 250 \, \& \, (t[1]==1 \, \| \, t[2]==1 \, \| \, t[3]==1)$\\
                                & & cycle has elapsed since last data acquisition.        & \\ \hline
    \multirow{9}{*}{\textbf{G3}}
                                & & Whenever the sensornodes' controller grants permission  & $A\square \,sensornode.ready \,$ \& $\, sensornode.exe==true$\\
                                & \textbf{P4} &to execute and its on high risk, $t\_high$ schedulers' cycle may        & $\& \, sn\_status==high \,$ imply $\, t==t\_high$\\
                                & & have passed since the last data acquisition.                & \\ 
    
                                & & Whenever the sensornodes' controller grants permission& $A\square \,sensornode.ready \,$ \& $\, sensornode.exe==true$\\
                                & \textbf{P5} &to execute and its on moderate risk, $t\_mod$ schedulers' cycles & \& $\, sn\_status==moderate \,$ imply $\, t==t\_mod$\\
                                & &may have passed since the last data acquisition.                & \\
                                & & Whenever the sensornodes' controller grants permission& $A\square \,sensornode.ready \,$ \& $\, sensornode.exe==true$\\
                                & \textbf{P6} &to execute and its on low risk, $t\_low$ schedulers' cycles may            & \& $\, sn\_status==low \,$ imply $\, t==t\_low$\\
                                & & have passed since the last data acquisition.                & \\ \hline                                                 
    \multirow{3}{*}{\textbf{G2}}
                                & & Whenever a sensor node has collected data,                  & \\
                                & \textbf{P7} & the bodyhub will eventually process it.                     & $sensornode.collected \to \, bodyhub.processed $ \\  \hline
    \multirow{3}{*}{\textbf{G4}}
                                & & Whether the sensornode has collected some data,             & \\
                                & \textbf{P8} & eventually the bodyhub will persist it.                      & $sensornode.collected \to bodyhub.persisted$ \\  \hline
    \multirow{3}{*}{\textbf{G5}}
                                & & Whenever a sensor node has sent data,                       & $sensornode.sent \to \,(bodyhub.processed \, \&$ \\
                                & \textbf{P9} & the bodyhub will eventually process             & $(bodyhub.received==low \,$ $\|$ $\, bodyhub.received==moderate \, \|$ \\ 
                                & & low, moderate or high data.                                 & $bodyhub.received==high)$ \\
                                & & Whether the bodyhub has processed some data, it             & \\
                                & \textbf{P10} & eventually will detect a new patient health status.          & $bodyhub.processed \to bodyhub.detected$ \\  \hline
    \end{tabular}
\end{table*}
\normalsize

In our approach, we specify temporal logic formulae to verify the satisfiability of the goals modeled in the CGM (cf. Figure~\ref{fig:goaltree}). The CGM root goal explicitly elicits the actor's main goal that once satisfied assures the system's correct behavior. Fulfilling hard goals requires its refinements by means of its AND- and OR-refinements into goals or tasks are satisfied. Taking into account the modeling strategy, the task's behavior can be verified through reachability properties as the timed automata locations represent the progress~and achievement of the task's behavior. For example, T1.1 (Collect sensor data) with $A\lozenge \, sensornode.collected $ meaning ``the sensor node will eventually collect sensor data'', also T2.2 (Detect patient health status) with $A\lozenge \, bodyhub.detected$ as ``the bodyhub will eventually detect the patient's status'', and so on. 

On the other hand, sequences of states comprising temporal relations need to be addressed in order to fulfill the CGM goals. These are achieved basically by combining task properties (\emph{reachable locations}) in the UPPAAL models with invariants or path-like formulae. Table~\ref{tab:properties} describes the sufficient TCTL-like specifications to assure the correct system behavior of the BSN by means of the satisfiability of the goals modeled in the CGM (cf.  Figure~\ref{fig:goaltree}).

The safety property P1 is common to distributed systems and assures that the BSN model is deadlock free. The fairness property P2 assures that in a scheduling cycle all modules will be executed. 

Regarding P3, it is noticed that goal verification is not trivial when it does not have a direct relation to task decompositions, specially when non-functional goals contributes to it, which is the case of the root goal G1 that demands the emergency detection within 250 ms (cf. Figure~\ref{fig:goaltree}). This is addressed in UPPAAL through an \emph{observer} automaton that records the time taken when high risk is acknowledged at the sensor node to its proper detection on the BodyHub.

To fulfill G3, the properties P4, P5, and P6 shall be assured as they represent the controller behavior in task T3 operationalization with respect to the frequency at which sensors data will be collected. Finally, the goals G4 and G5 are satisfied by assuring their means-end task executions. In particular, G5 guarantees through property P9 that the sensor data range follows the BSN operationalization data accordingly, that is, low, moderate or high. This is important to assure that the model does not have any data sent outside the range recognized by the BSN system. G2 is the fulfillment of property P7, which merely synthesizes the fulfillment of goals G4 and G5.

\subsection{From Goals to the Data Mining Process}\label{sec:3-2}

Over the past years, artificial intelligence techniques, specifically machine learning, have been adopted to deal with the difficulty in predicting context conditions at runtime~\cite{ShariflooMQBP16}. Knowing that the use of such techniques at runtime poses a risk to the performance of SAS, we intend to provide assurance in terms of requirements compliance through an off-line model verification, that is, the verification is not directly connected or controlled by the running system~\cite{Weyns2016PerpetualAF}, and complement it on-line, through the use of data mining techniques over data generated by the running system or prototype. 

As object of the mining process, we apply the concept of \emph{transfer learning}, that is, the process of using other sources to provide cheaper samples for accelerating model learning~\cite{TransferLearningSurvey}, for the provision of legitimate data that will cope with the context analysis process. The data mining process enables us to isolate the component behaviors that need to be analyzed. The scope of analysis scales as we go up in the CGM treelike structure, that is, encompassing more elements in a database record and embracing all TCTL properties. By these means, the outcome of the data mining process provides evidence for the assurance check in our methodology. Additionally, the data mining process for relevant context conditions (i.e., where the SAS operates) does not necessarily need to make a combinatorial exploration to every possible system state. Instead, it analyzes the relevant properties, following from those formally specified in UPPAAL, and the CGM tasks' contexts under operation, which speeds up the learning process. 
As such, the enrichment of the goal model with causal relationship between context and the non-functional requirements fulfillment supports the anticipation of adaptation strategies and potentially mitigates runtime uncertainty. To bypass both time and space limitations, we aim at merging the context discovery potential by means of artificial intelligence over monitored data, typical of runtime approaches, with the perks of having a robust modeling process at design time. The core of our idea with respect to applying data mining techniques relies on the mining and analysis of the impact that contexts might have on the satisfaction of real-time properties for SAS.

The classification routine is the main technique we use in the scope of our work. It is a data mining technique that tackles the problem of identifying to which set of categories an observed fact belongs. It is done based on a training set of data containing observations (or instances) whose category membership is known, that is, it is an instance of supervised learning~\cite{Alpaydin2014}. More specifically, we apply two classification methods in our approach: 
(i) JRip~\cite{Cohen}, that creates rules for every class in the training set and then prunes these rules. The discovered knowledge in this class of algorithm is represented as IF-THEN prediction rules and are specially useful to define the operational thresholds of some resources.
(ii) The J48 classifier algorithm, which implements a Decision Tree used to support the decision making process using the depth-first strategy~\cite{Quinlan}. In our method, we benefit from the decision trees by showing: (i) if the behavior of the actual SAS conforms to the verified properties and (2) how the different contexts of operation are combined to achieve a given goal focusing on a target property value. 
In the case of a non-nominal classification, that is, when the target knowledge or prediction concerns to a numeric attribute, it is possible to replace the decision tree for other kinds of model trees, for instance, ones that work with linear regression such as M5P~\cite{M5P}.

In a nutshell, we propose an analysis process using the aforementioned algorithms (J48 and JRip) to unveil operational contexts of the SAS that might influence the satisfaction of non-functional requirements and real-time properties and, at the same time, quantify such impacts on tasks and/or goals. Based on the CGM structure (cf. Figure~\ref{fig:goaltree}) combined with the operationalization values of sensed information, we traverse the goal-tree visiting each CGM node (goal or task) and verify how the CGM nodes and the properties related to such nodes (\emph{vide} Table \ref{tab:properties}) corroborate the design-time verification in a runtime environment. Such verification is supported by the data mining and analysis of the log generated by a SAS, in our case the BSN prototype in OpenDaVinci with its CGM goals (see Figure~\ref{fig:goaltree}).

\subsubsection{The Data Mining Process}

We provide a fine-grained presentation of our data mining process with the algorithmic steps of Algorithm \ref{func:algo}. The algorithm starts by traversing the CGM using a post-order depth-first search (DFS) in line~\ref{algline:TreeTraversal}. After running the SAS, for each node of the CGM, the portion of the log dataset, that is, particularly the variables related to the node's subtree (line~\ref{algline:subtreeselect}), is collected and processed, more specifically those variables encompassed by that node's subtree. In lines~\ref{algline:getproperties} and~\ref{algline:getctxt} we get, respectively, the properties and the contexts of operation that are associated to that specific subtree. 
In possession of the dataset, we execute the JRip algorithm to extract the operationalization rules (line~\ref{algline:JRIP}), and we apply the J48 to display the combination of such rules with the context observed at runtime (line~\ref{algline:J48}). 
At last (line~\ref{algline:analysis}), the knowledge obtained from the data mining process is confronted with the UPPAAL properties previously defined and verified, allowing us to unveil any dissonance between the design/runtime model and the real execution. Moreover, the knowledge obtained by the process allows us to improve and fine tune the adaptation policies, maximizing a target attribute while observing the behavior of the whole system, either at design-time via simulation or at runtime. All this is possible through the identification of facts that are not noticeable by the model verification alone.

\begin{algorithm}[!h]
 \caption{DataMiningToSupportAssurance}
 \label{func:algo} 
\begin{algorithmic}[1]
\Require SubTree dataset, CGM cgm, RTModel properties \label{algline:syntax}
\Ensure RTContextualKnowledge

\State Analysis rules  $\leftarrow$ NULL
\State Analysis decision\_tree  $\leftarrow$ NULL
\State Context ctxt  $\leftarrow$ NULL
\ForAll{n \textnormal{\textbf{in}} cgm \label{algline:TreeTraversal}}
\State {SubTree dataset $\leftarrow$ operational\_data.OpenDaVINCI(n)\label{algline:subtreeselect}}
\State{RTModel properties $\leftarrow$ cgm.getproperties(n)\label{algline:getproperties}}
\State ctxt.push(dataset.getContext(n))\label{algline:getctxt}
\State rules.push(dataset.JRip())\label{algline:JRIP}
\State decision\_tree.push(dataset.J48())\label{algline:J48}
\EndFor
\ForAll{n \textnormal{\textbf{in}} cgm \label{algline:TreeTraversal2}}
\State RTContextualKnowledge $\leftarrow$ analyse(n.rules, n.decision\_tree, n.properties, n.ctxt)\label{algline:analysis}
\EndFor
\end{algorithmic}
\end{algorithm}

\begin{figure}[!h]
    \centering
    \includegraphics[width=0.85\columnwidth]{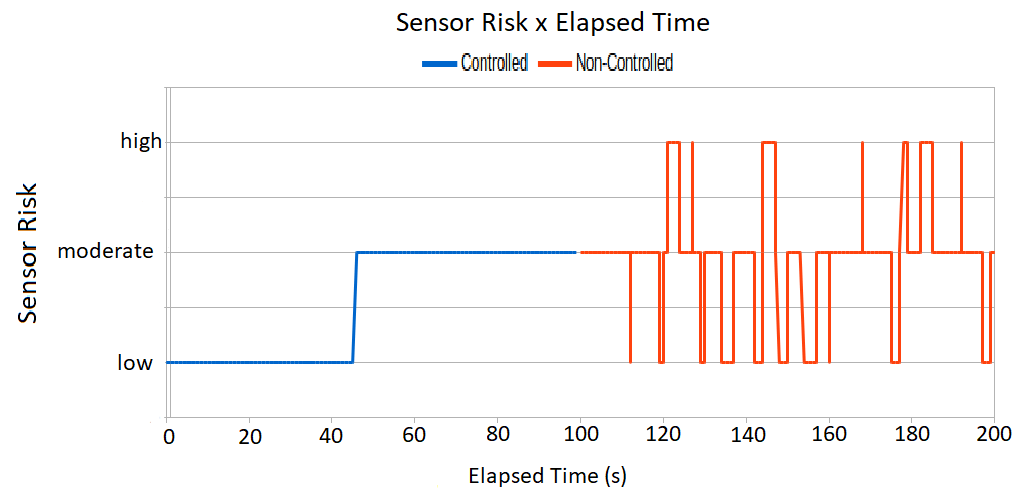}
    \caption{Behavior with and without the context \emph{controller}.}
    \label{fig:contcontext}
\end{figure}

To illustrate how the data mining provides evidence for assurance, let us explain the process in a practical sense using the BSN as an example. Most of mobile health care applications depend upon batteries for several services. Therefore, the predictability of the duration of a battery cycle is paramount in medical applications to guarantee that the devices' availability are always observed. 
Considering that the probability of triggering an emergency signal is directly proportional to the patient health risk and to guarantee the safety of a high-risk patient, it requires a continuous monitoring of the patient's status, that is, a high sampling frequency of the patient vital signs. 
On the other hand, for a low-risk patient, a sporadic monitoring is sufficient to guarantee the patient's safety. Therefore, the higher the monitoring frequency the higher the energy consumption of the device processing the sensed vital data. Our BSN's controller is responsible for managing the frequency of sampling rate of the patient's vital signs, according to her health risk status.  
For instance, when the controller context is present (i.e., the controller is activated), the period of data acquisition could be retarded for low and moderate risks in a factor of ten and five times, respectively. The importance of the controller policy can be seen in Figure~\ref{fig:contcontext}, which shows how frequently the BSN sensors in our prototype are required to perform, that is, obtaining the patient's vital signs and categorizing the patient's health risk status, under the presence or absence of the controller. So the issue lies on defining suitable parameters for the controller policy that is supported by an efficient energy consumption, but on the other hand still satisfies the verified properties of the SAS.

\begin{figure}[!h]
    \centering
    \includegraphics[width=0.8\columnwidth]{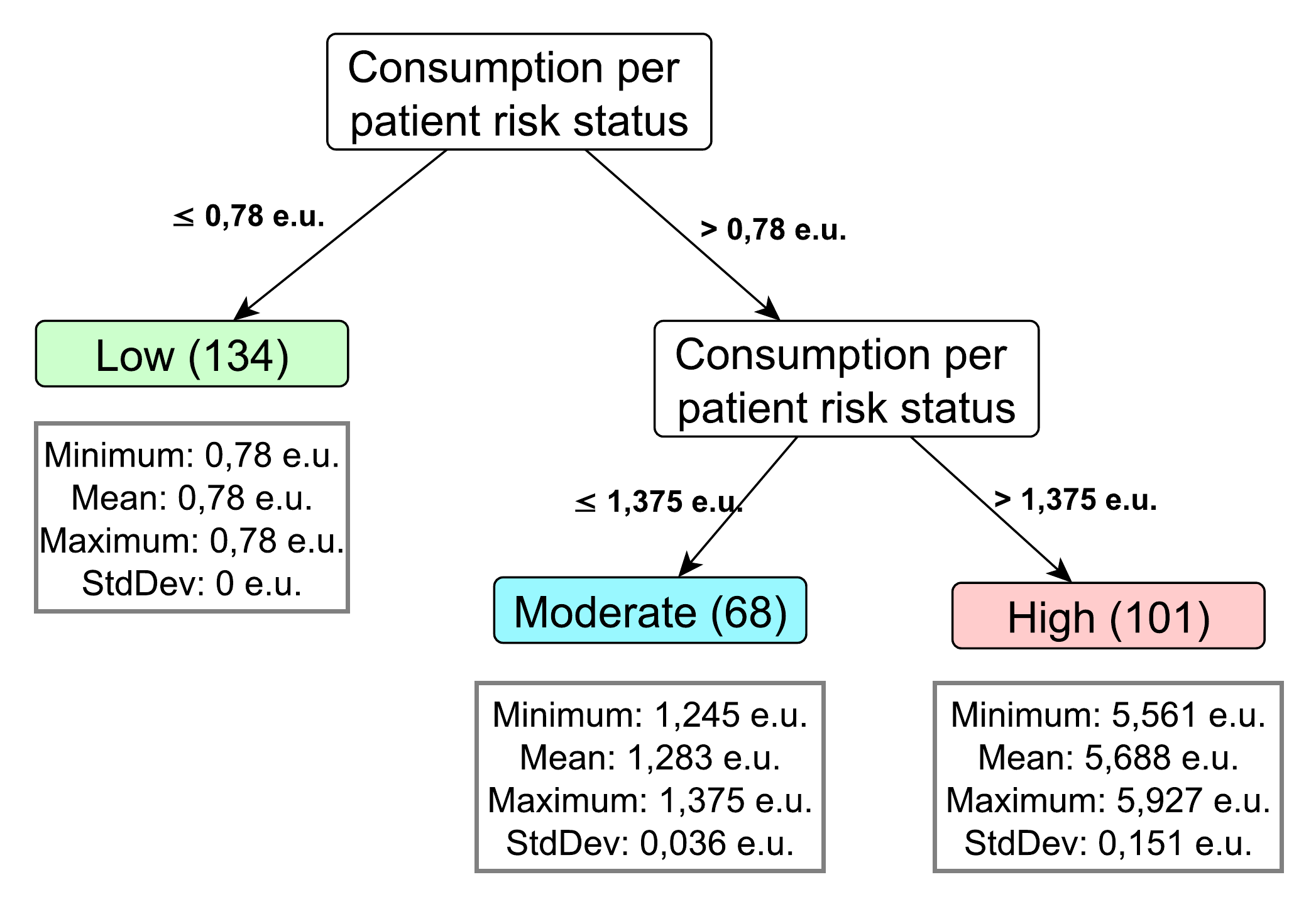}
    \caption{Battery consumption per health risk status.}
    \label{fig:BatteryPatientTree}
\end{figure}

The knowledge about the overall energy consumption in face of a controller's context variability (context C2 in the subtree of goal G3, Figure~\ref{fig:goaltree}), can be enhanced by our data mining process to find out a precise outcome that could not be obtained merely by a model checking process. For example, as an outcome of our data mining process, Figure \ref{fig:BatteryPatientTree} presents a decision tree of the energy consumption in energy units (e.u.) per health risk status of a patient on the BSN prototype. 
Such values in the decision tree nodes were obtained from monitoring the battery consumption from the BSN prototype implemented in OpenDaVINCI and, aiming at the normalization of the analysis for the user profile simulated. As such, our data mining process helps on quantifying and classifying the battery consumption per health risk status of a patient in the controlled context mode. 

Therefore, after learning from our data mining process, it is possible to draw a relationship between consumption per risk status for a specific user profile. Through the rules generated by JRip and the decision trees provided by J48, it is possible to define ranges of operation in which energy will be drained at a predictable rate. However, one should note there were no checked properties regarding the energy consumption in the CGM, nor in the properties verified in UPPAAL. Nevertheless, our data mining process was able to provide means to support the assurance process regarding the controller mode of operation. For example, since the controller mode of operation plays an important role on the energy consumption in our BSN, our data mining process could support the creation of adaptation policies to dynamically adjust the sampling frequency. These policies are used by the sensor node's controller to optimize the energy consumption in a specific context (battery level) without violating its verified properties. 

In the next section, we further present the outcomes of the experimentation of our learning-based approach to support assurance for real-time SAS on the BSN prototype developed in OpenDaVINCI.
\section{EXPERIMENTS IN OPENDAVINCI}\label{sec:Experiments}

The BSN prototype was developed in OpenDaVINCI~\cite{ODV}, that stands for Open Source Development Architecture for Virtual, Networked, and Cyber-Physical System Infrastructures. OpenDaVINCI is ideal for networked cyber-physical applications, as it enables TCP, UDP, and serial port communication. The platform permits real-time scheduling for distributed software architectures. Working as a middleware responsible for data- or time-triggering software modules, it deals with message distribution on publish/subscribe communication architecture. Moreover, it supports real-time operations under a real-time Linux system. Therefore, it became a natural choice to the development of the prototyped real-time BSN.

\subsection{Metrics}

We evaluate our approach by means of a Goal-Question-Metric (GQM) methodology~\cite{GQM}. The questions that are relevant to the evaluation of the present work are divided into three major experiments that analyze: (i) the impact of the number of BSN sensor nodes on the satisfaction of real-time constraints, (ii) the impact of patients health risk diagnosis strategies on the satisfaction of real-time constraints, and (iii) the impact of the controller context on the energy consumption of the BSN. Table~\ref{GQMPlan} details these questions. 

We have adopted two different time metrics to illustrate our results. The first one is $T_{SN}$, which represents the time difference between two consecutive measurements of a given sensor node. In terms of TCTL specification, the metric is described as $T_{SN} = t_{sensornode[k].collected} - t_{sensornode[k+1].collected}$. The second time metric is $T_{ED}$, referring to the time taken to detect an emergency after a patient's health status is identified as high risk. Transcribing to TCTL, the $T_{ED}$ metric is represented as $T_{ED} = t_{bodyhub.processed} - t_{sensornode.collected}$, that is, the time difference between the sending of measured data by a sensor node and the processing of the data by the BodyHub. 

\begin{table}[t]
\caption{GQM plan}
\label{GQMPlan}
\vspace{-1.5em}
\footnotesize
\begin{tabular}{|p{0.29\textwidth}|p{.15\textwidth}|}
\multicolumn{1}{l}{}\\\hline
\multicolumn{2}{|c|}{\textbf{Goal 1: Property Refinement through Data Mining Process}}\\\hline
\multicolumn{1}{|c|}{\texttt{Question}}& \multicolumn{1}{c|}{\texttt{Metric}}\\\hline

1\hspace{0.4cm}\texttt{Are the scheduling period (property P2) and emergency detection time constraint (property P3) respected while varying operational contexts such as the number of active sensors?} & \texttt{$T_{SN}$(s), $T_{ED}$(s).}\\\hline

2\hspace{0.4cm}\texttt{How does the data mining process support the maximization of the trustworthiness of a measured data while respecting a tight scheduling window?}& \texttt{$T_{ED}$(s), Paired t-test.}\\\hline

3\hspace{0.4cm}\texttt{How does the learning method assist the unveiling of sensitive runtime aspects and guide the fine tuning of the BSN's adaptation policies?} & \texttt{Average battery consumption (\%) for patient risk status.} \\\hline

\end{tabular}
\end{table}

\subsection{Setup}

The presented experiments were executed under the following configuration: CPU 4x Intel(R) Core(TM) i7-5500U CPU @2.40GHz, 8075MB RAM, Ubuntu 16.04.3 LTS, Kernel Linux 4.4.86-rt99 (x86\_64) with SMP PREEMPT RT, GNU C Compiler version 5.4.0 20160609, hard drive ATA Corsair Force LE. 

We have divided the evaluation of our work into three major experiments, each one related to a question described in Table \ref{GQMPlan}. All the artifacts related to these experiments are available at GitHub\footnote{\url{https://github.com/rdinizcal/SEAMS18}}.

We should note that sensors are prone to failure so that we applied in our experiments two different strategies to confirm the vital sign status sent by the sensors. In the first strategy, namely \emph{3 of last 5}, the system verifies whether the same vital sign status is confirmed in at least three out of the last five reads of the same sensor. Only in this case the data is sent to the central unit. In the second strategy, namely \emph{Replication}, each sensor node comprises 5 redundant sensors and the system takes the measurement of the majority of the replicated sensors as a valid reading.

\subsection{Results}

\paragraph{\textbf{Q1: Are the scheduling period (property P2) and emergency detection time constraint (property P3) respected while varying operational contexts such as the number of active sensors?}}

First of all we need to characterize the time constraints related to the BSN's real-time properties. The first time constraint we analyze is the scheduling cycle period, whose violation potentially leads to the violation of the fairness property of the system (P2). According to this property, all the sensor nodes have to be executed within the time window of 100\emph{ms}, which is the default scheduling period of OpenDaVINCI working at 10Hz. We say that the scheduling period constraint is respected if the time difference between two consecutive measurements of a given sensor node is less than (or equal) to 100\emph{ms} ($T_{SN} \leq 100 ms$). 
The second time constraint refers to the property P3 presented in Table \ref{tab:properties}. This property is related to the goal G1, displayed in Figure \ref{fig:goaltree}, concerning the time threshold of 250\emph{ms} in which an emergency shall be detected after a patient's health status is identified as high risk. In order to satisfy this constraint, the time difference between the sending of a measurement by a sensor node and the processing of such reading by the BodyHub is less than (or equal to) 250\emph{ms} ($T_{ED} \leq 250 ms$). We should also note that the experiments have been made with different numbers of sensor nodes (1, 5, 10, and 20 sensor nodes) to investigate the consequences when increasing them. As such, each sensor node would contain different number of sensors, depending on the confirmation strategy: in the \emph{3 of last 5} it would contain only one sensor, while in the \emph{Replication} it would contain 5 redundant sensors.

\begin{figure}[!h]
\hspace*{-0.8 cm}
    \centering
    \includegraphics[width=1.2\columnwidth]{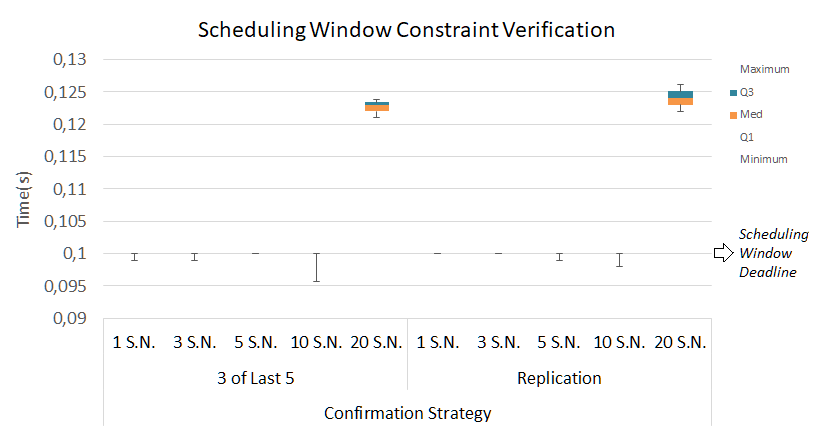}
    \caption{Scheduling window satisfaction for different contexts ($T_{SN} \leq 100 ms$).}
    \label{fig:SchedulingWindow}
\end{figure}

We have verified for the first time constraint that all the mean values are below 100\emph{ms} either using the \emph{3-of-last-5} and \emph{Replication} strategies, except of the scenarios with 20 sensor nodes, in which the $T_{SN}$ values for both strategies are around 123\emph{ms}. To satisfy all configurations and time constraints, the upper bound limit is determined by the execution with 20 sensor nodes and applying \emph{Replication} strategy, that is, approximately 126\emph{ms}. 
Therefore, the scheduling period constraint (P2) was not fully satisfied for any configuration as it is depicted in Figure~\ref{fig:SchedulingWindow}. To summarize, the increase of the number of sensor nodes is a sensitive factor for the satisfaction of the real-time constraints. The execution of the system with more than 20 sensor nodes hinders the monitoring/processing of the patient's vital signs and is potentially the cause of data loss, jeopardizing the integrity and availability of the BSN during the scheduling routine.

\begin{figure}[!h]
\hspace*{-0.8 cm}
    \centering
    \includegraphics[width=1.2\columnwidth]{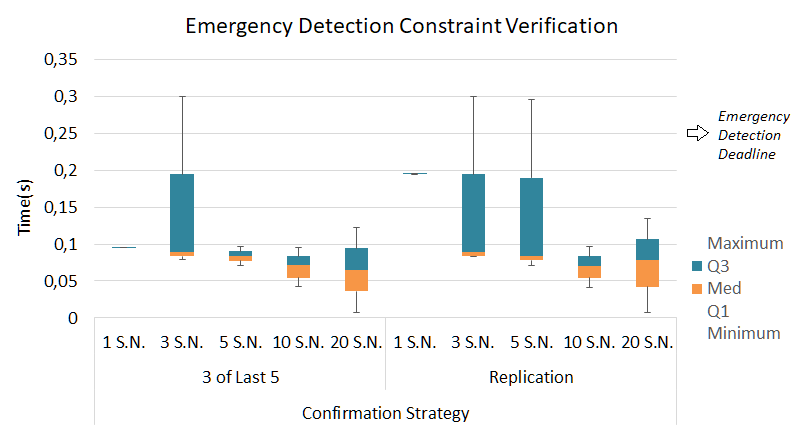}
    \caption{Emergency detection satisfaction for different contexts ($T_{ED} \leq 250 ms$).}
    \label{fig:EmergencyDetection}
\end{figure}

For the second time constraint that refers to the emergency detection, we have noted that the property P3 was respected in every scenario with exception of some outliers observed in scenarios with 3 sensor nodes running with \emph{3-of-last-5} strategy, as well as 3 and 5 sensor nodes using \emph{Replication}. The evaluated scenarios and their respective $T_{ED}$ values are shown in Figure \ref{fig:EmergencyDetection}.

The data mining analysis has exposed some BSN aspects that, from a model checking perspective, do not impact the system behavior, but at runtime pose a threat to the fulfillment of real-time requirements. The method allowed us to spot the sources of unexpected changes that can only be verified through runtime data, such as the variability in the amount of active sensors, confirmation strategy, or the controller mode. The data mining process also pointed out that part of the measurements are received by the BodyHub \textit{after} the scheduling cycle period closes. Although the OpenDaVINCI implementation was able to store the messages in a buffer and process all of them afterwards, validating property P9, we cannot guarantee that the same would happen in a real-world scenario. We can only infer that as we employ more sensor nodes, more measurements tend to violate the scheduling window, potentially putting at risk the fairness property of the system (P2) as well as P9.

\paragraph{\textbf{Q2: How does the data mining process support the maximization of the trustworthiness of a measured data while respecting a tight scheduling window?}}

In real-world scenario, sensors are considered failure-prone. This experiment aims at verifying how sensors data validation strategies scale in a system with tight real-time constraints and how the data mining can assist the process. We have applied a pairwise t-test to compare two population means ($T_{ED})$ where we have two samples in which observations in one sample (\emph{3 of last 5} strategy) can be paired with observations in the other sample (\emph{Replication strategy}). We can use the results from our sample emergency detection to draw conclusions about the impact of changing the strategy in general.

To calculate the confidence interval for the true mean difference, at a 95\% confidence interval the true mean difference is: $ -0.02902\pm(2.776 \times 0.01939) = -0.02902\pm0.05382 $. The result confirms that the difference in strategies $T_{ED}$ is not statistically significant, since the interval (-0.08284, 0.02484) includes 0.

\begin{figure}[!htb]
    \centering
    \includegraphics[width=\columnwidth]{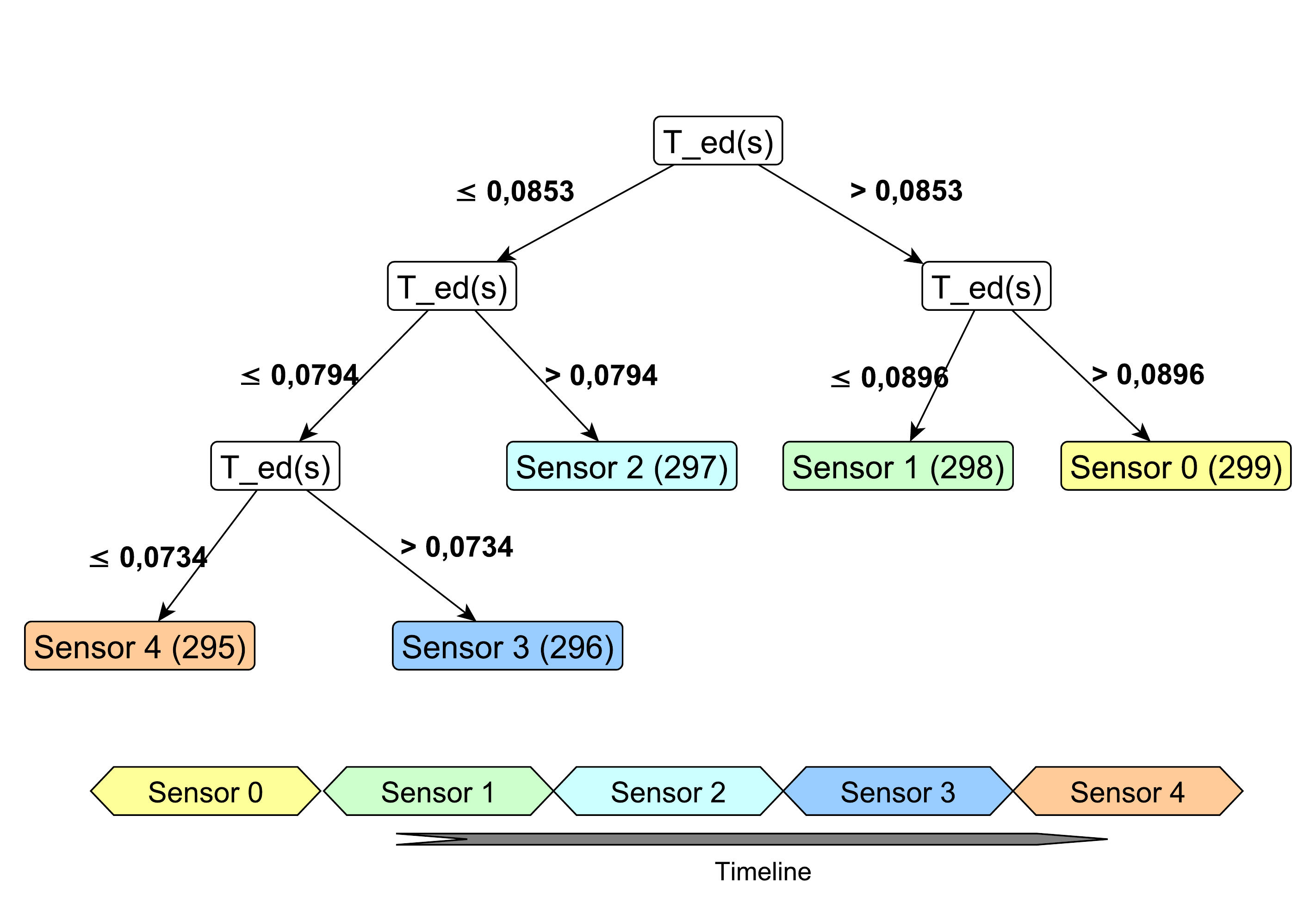}
    \caption{Sensor node queue and their processing times.}
    \label{fig:QueueTree}
\end{figure}

The J48 method with its generated decision trees assisted us in defining the processing time of each sensor while sensors with higher computation times can be executed later in the scheduling sequence. Figure~\ref{fig:QueueTree} shows how the execution queue of the sensors can be sorted based on the processing period. The decision tree is useful to assist the debugging process of a SAS, finding, for instance, the modules that are potential sources of failures. In the BSN case, it helped us to spot bottlenecks in the patient monitoring and possible data loss from sensors. This enables the development of a more dependable SAS by fine tuning and hence obtaining robust adaptation rules.

\paragraph{\textbf{Q3: How does the learning method assist the unveiling of sensitive runtime aspects and guide the fine tuning of the BSN's adaptation policies?}}

\begin{figure}[!h]
    \centering
    \includegraphics[width=\columnwidth]{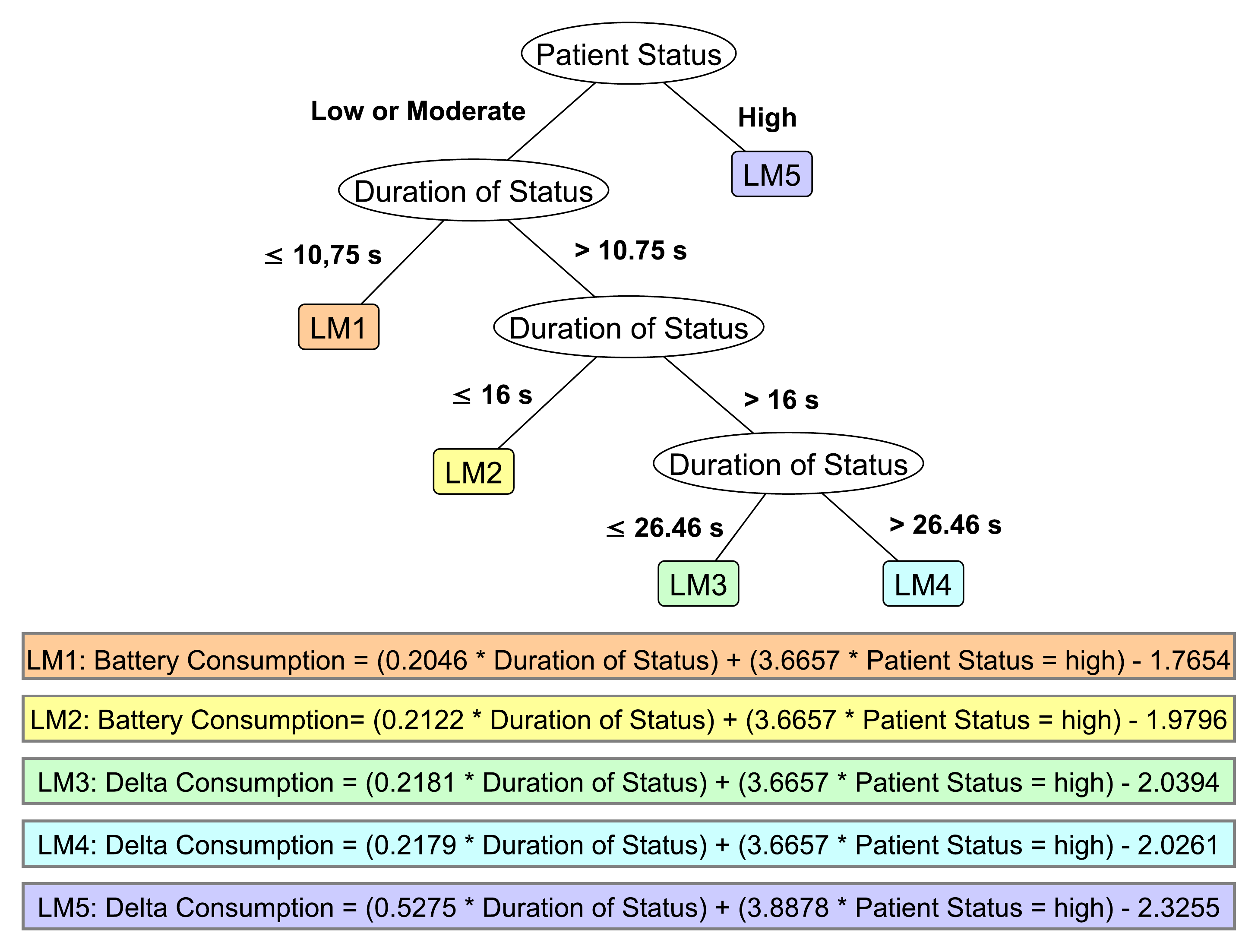}
    \caption{Multivariate multiple regression technique to enhance adaptation policy.}
    \label{fig:M5P}
\end{figure}

Defining adaptation policies at design time with respect to aspects that can only be known at runtime is a challenging task for engineering a SAS. Since the sampling rate variable is highly dependent on the user profile (e.g., health status), the data mining process stands out on fine tuning the related policies assisted by runtime data. By introducing a multivariate multiple regression technique into our process, we are able to estimate the duration in which a patient stays in a health risk status before it changes to another one (see Figure~\ref{fig:M5P}). Moreover, it enables the prediction of energy consumption based on the current patient status and duration in such a status. As we have mentioned before, the patient health risk state is directly related to the battery consumption. Therefore, by merging the knowledge obtained via data mining, we could predict for how long a patient can be reliably monitored given the patient's current health risk and the remaining battery charge.

Based on the output of the data mining process, we devised a dynamic adaptation policy for the controlled mode to balance the trade-off between the energy consumption and safety assurance. The battery charge was divided into three categories: good (50\%-100\%), medium (15\%-49\%), and critical ($<$15\%). Basically, for a conservative monitoring policy, the system keeps track of the minimum duration in which a patient's health risk status remains unchanged for each status (low, moderate and high). Such a period will be the parameter to adjust the sampling rate in each situation. When the battery charge reaches the medium level, the adaptation planner takes the minimum duration in which a patient stayed in a given risk during the measurement lifetime, and set this duration as the sampling rate for that risk status. 
For instance, if an individual patient stays on average 3 hours in moderate risk, but the minimum duration measured for such status was 30 minutes, the latter will be the new sampling rate for this risk state. This policy will minimize the energy consumption and at the same time maintain the assurance of the real-time constraints. Finally, when the system notices that it will not be able to guarantee the next measurement due to insufficient charge, that is, the sampling period is greater than the estimated working time for current charge level, it enters in an \emph{energy saving profile} and adapts the replication strategy. Instead of taking the majority of five readings, it shuts down two sensors and take the best of three measurements in order to save even more battery.

\begin{figure}[!h]
    \centering
    \includegraphics[width=\columnwidth]{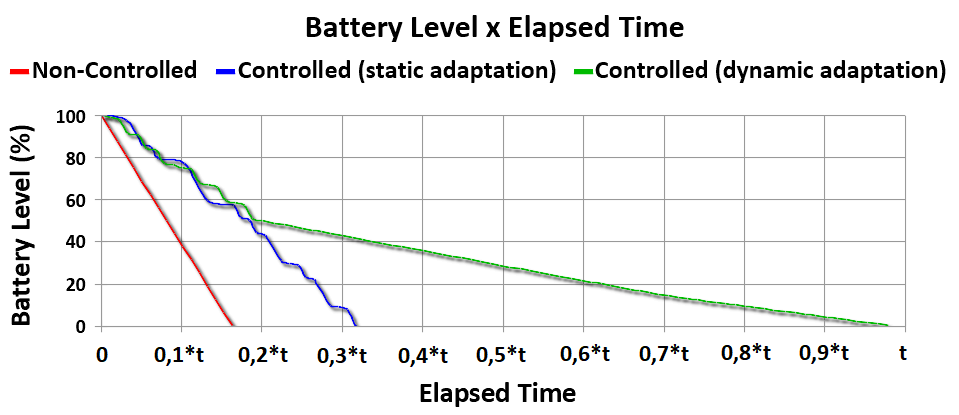}
    \caption{Battery consumption w.r.t. the controller context.}
    \label{fig:BatteryConsumption}
\end{figure}

Figure \ref{fig:BatteryConsumption} shows the progression of BSN's energy consumption over time for a patient's monitoring considering (i) a non-controlled scenario, (ii) a controlled scenario with a static adaptive policy, and (iii) a controlled scenario with a dynamic adaptive policy, that is, the default policy enhanced with the knowledge obtained by the data mining process. 
We have noticed that the battery, in the policy supported by the learning process, lasts over three times longer in comparison to the other contexts. During this time, we were able to reliably identify all the patients' statuses and their variations without violating the real-time constraints. 

Closing the feedback loop, the UPPAAL model was updated as well to confirm whether the properties described in Table~\ref{tab:properties} still hold after the changes introduced in the adaptation policy. The obtained results were encouraging since we were able to identify some runtime aspects that are directly related to the satisfaction of the real-time constraints and to the optimization of quality attributes such as energy efficiency. Our method stood out in identifying such aspects and manage them to improve the dependability of a system in a cost-effective manner. Previously, such knowledge could only be identified at runtime via a reactive approach. With our method, we have access to anticipated information that assists the (i) validation of design-time properties, (ii) development and refinement of adequate adaptation policies, and (iii) the assurance in decision making process, even at design time through simulating a SAS prototype.

\subsection{Threats to validity}

\textbf{Construct validity} -- Our input data relied on a reported sound case study (BSN) and its published and available data. In addition, our process aligns goals, model checking, prototyping and data mining following a sound procedure. Despite all the care we took to avoid the generation of unrealistic data, further study must be done to verify the applicability in a real-world scenario. 

\noindent\textbf{Internal validity} -- Our approach showed itself efficient to adequately evaluate our approach. However, unveiling all the contexts involved in a system's operation is inherently NP-complete, which could represent a threat to the overall assurance of the system.

\noindent\textbf{External validity} -- Although our approach is not tailored to be domain specific, we do reckon the limitation of the evaluation since it was applied in the specific case of the BSN. Further evaluation must be performed to evaluate the generalization capability.%
\section{RELATED WORK}\label{sec:RelatedWork}

Among the approaches that aim at assurances for self-adaptive systems through design-time verification and validation, C{\'{a}}mara and Lemos~\cite{CamaraL12} define an approach that relies on the notion of stimulation and probabilistic model-checking to provide levels of confidence regarding service delivery, with focus on the resilience property. Still on model-checking, de la Iglesia~and Weyns~\cite{IglesiaW13} extend an agent-based mobile learning application with a self-adaptation layer. The authors also developed a set of formally specified MAPE-K templates that encode design expertise for a family of self-adaptive systems~\cite{IglesiaW15}. In the domain of advanced distributed embedded real-time systems, Giese \emph{et al.}~\cite{GieseS13} propose MechatronicUML, a model-driven development approach which supports the modeling and verification of safety guarantees for SAS at runtime. Calinescu et al.~\cite{CalinescuGHIKW17} have recently proposed the end-to-end ENTRUST methodology through systematic stages to provide assurance evidence, cases and arguments for the controller platform at design and runtime for SAS. 
In our work we increment the assurance processes with model checking by means of a learning-based approach to verify and validate real-time properties while accounting for context variability, including adverse conditions, at runtime.

Regarding runtime models in SAS, Chen \emph{et al.}~\cite{ChenPYNZ14} propose the combination of requirements-driven self-adaptation and architecture-based self-adaptation to reconfigure component-based architecture models using incremental and generative model transformations for complex architectural adaptations.  
Vrbaski \emph{et al.}~\cite{VrbaskiMPA12} propose a work that leverages goal models as runtime entities and integrates them into modeling, simulation, and execution environment of context-aware systems. As a complement to the requirements-driven self-adaptation approach proposed by Qian \emph{et al.}~\cite{QianPCMWZ15}, that combines goal
reasoning and case-based reasoning, our approach explores how the combination of contextual variables, systems' configurations and non-functional requirements affects the selection of the adaptation solution. In our approach, we similarly rely on the modeling structure of the contextual model. 
But most of all, we rely on a feedback loop to keep a verification model always up-to-date, accounting for the real-time properties, as well as others, aligned with the dynamic aspects of the system. 

In the field of machine learning and data analysis to support the modeling and adaptation of self-adaptive systems, Sharifloo \emph{et al.}~\cite{ShariflooMQBP16} propose a solution for design-time uncertainty, particularly in the realm of Dynamic Software Product Lines (DSPL), by proposing a feedback approach through an adaptive system model that combines learning of adaptation rules with evolution of the DSPL configuration space. Knauss \emph{et al.}~\cite{AlessiaACon} propose a framework to provide adaptation of contextual requirements at runtime. Our work, on the other hand, supports the provision of evidence for SAS assurance still at design time. To this purpose, we consider the impact of contexts combinations in real-time constraints and dependability attributes, and use such knowledge to improve the system's adaptation. Aiming at the enhancement of the learning process, Jamshidi \emph{et al.}~\cite{JamshidiVKSK17} define a cost model that transform the traditional view of model learning into a multi-objective problem, considering model accuracy and measurements effort. 
Similar to our approach, their objective is to apply the concept of \emph{transfer learning} to improve model predictions in SAS. Still in the domain of large scale distributed systems, Schmid \emph{et al.}~\cite{SchmidGPB17}, tackle the difficulty in developing a complete and accurate model for SAS at design time by proposing a method where the system model consists only of the essential input and output parameter. 
Our method also benefits from the ability of classification methods particularly by (1) discovering hidden patterns of operation in presence of different contexts of execution, and (2) supporting with evidences that the model-checked properties hold for the running system.

Regarding the assurance of real-time properties, Zeller and Prehofer~\cite{ZellerP12} deepen the study of time constraints for runtime adaptation analyzing two approaches for finding solutions in the resulting search space for adaptations, one based on planning algorithms and the other based on constraint solving. In our work we verify the applicability of data mining techniques to assist us in defining  time constraints without sacrificing the performance in the contexts of operation. For systems with strict time constraints but where accuracy is not a major concern, statistical model checking could be an alternative~\cite{YounesS06, LegayDB10}. Initial research that uses statistical techniques at runtime for providing guarantees in self-adaptive systems is reported by Weyns and Iftikhar in~\cite{WeynsI16}. In our work, we go one step further on exploring context conditions and their implications on the real-time properties through the data mining, since we believe in domains such as the BSN, vital and accurate information require a more thorough perspective of analysis to make more evident whether the SAS do hold the properties even under adverse conditions.%
\section{CONCLUSION AND FUTURE WORK}\label{sec:Conclusion}

In this paper, we proposed an integrated method that applies the concepts of data mining and analysis over the runtime data of a SAS, that supports the formulation of hypothesis concerning the impact of context variability in non-functional requirements and time constraints. Using such analysis to feedback the approach allows us to validate and refine the properties collected at design time via a model verification process. Hence, we are able to reduce the gap between the design- and runtime models, reducing the uncertainty in the adaptation process. In our evaluation based on the published information of the BSN, we were able to model the variability observed in sensed data and perceive a significant impact on the outcomes with respect to real-time constraints, sensor nodes scheduling, controllers actuation, and energy profiling.
For future work, we plan to provide means to seamlessly integrate the steps of our approach as well as to explore unsupervised data mining approaches to be able of dealing with more complex contexts variations.%

\begin{acks}
This work has been partially funded by CAPES/PROCAD under Grant No. 183794, and Arthur Rodrigues' CAPES/DS scholarship.
\end{acks}

\bibliographystyle{ACM-Reference-Format}
\bibliography{acmart} 

\end{document}